\documentstyle[sprocl]{article}
\begin{document}
\title{SLOWING DOWN AND SPEEDING UP PSR'S
PERIODS:\\A SHAPIRO TELESCOPE TRACING DARK MATTER}

\author{D. Fargion, R. Conversano}

\address{Physics Dept. Rome University 1 and
INFN,\\ Piazzale A.Moro 2, 00185 Rome, Italy}

 \maketitle

\begin{abstract}
Dark matter influence globally galactic Keplerian motions and
individually, by microlensing, star's acromatic luminosities.
Moreover gravity slows down time as probed by gravitational
redshift. Therefore pulsar's (PSR's) periods may record, by
slowing and speeding up, any Dark MACHO crossing along the
line-of-sight. This phase delay, a Shapiro Phase Delay, is a
gravitational integral one and has been probed since 1964 on
planets by radar echoes. We discovered in PSRs catalog a few rare
PSRs (with negative period derivatives) consistent with speeding
up by Shapiro delay due to MACHOs. We propose to rediscover the
Shapiro effect on our solar system (for Juppiter, Mars, Sun) to
calibrate a new ``Shapiro Telescope Array'' made by collective
monitoring of the sample of PSRs along the ecliptic plane.
\end{abstract}

\bigskip
\bigskip
\section{Introduction}
\bigskip

Matter bends space-time. This is the reason for Newton's apple to
fall down and for planets and stars to follow Keplerian curves.
Gravity is the deformation of space and time. This double reality
is why massless photons are bent twice by Einstein Relativity than
just one time as predicted by a simplest Newtonian plus Special
Relativity model. This factor two is the base for 1919 light
deflection test and General Relativity triumph.\\
The light bending is the root for the microlensing techniques able
to reveal unseen dark MACHOs running in our galaxy. But time
itself is slowed down by gravity. The famous twin paradox might be
played as well as near and far a neutron star surface (neglecting
tidal-Riemmann forces). The clock beats are slowed down near
compact stars (NS, white dwarf, sun..) than in a far flat space-
time. Indeed gravitational redshift is an additional cornerstone
for General Relativity. The variable position of a wave-source
respect to a gravitational field is ``recorded'' by a phase shift
on the wave observed: a Shapiro Phase Delay. Indeed it has been
predicted on 1964 and observed by Shapiro himself with radar
echoes from Mercury and Venus, grazing the Sun. Because PSR's
period are on average stable ($<\dot{P}_{psr}> \;\leq
10^{\,-14}\;s\,s^{\,-1}$) we proposed [3,4,5] as timing candles to
scan all over the galaxy hunting for crossing Dark Matter along
the line of sight.

\bigskip
\section{The first evidences of Shapiro Phase Delay on PSR's}
\bigskip

Microlensing technique seems an unique powerful test because it is
inspecting million stars for acromatic luminosity magnifications:
how could a few hundred pulsar sample compete with such a huge
number? The reason lay on the extreme Shapiro Phase Shift
sensitivity versus the microlensing one. While microlensing
magnification is proportional to the inverse of the MACHO's impact
parameter $\sim 1\,/\,b_{min} = 1\,/\,u_{min} R_E$ (where
$u_{min}$ is adimensional and $R_E$ is the Einstein Radius),
Shapiro Phase Delay is made up by two main terms: a first
geometrical one, $\Delta t_{geo} = \frac{r_S}{4 c} \left(\,
\sqrt{u^2 + 4 \;} \,-\, u\,\right)^{\,2}$, which is due to deflect
and undeflected path and it is rapidly vanishing at large impact
values:  $\left. \Delta t_{geo} \right|_{u \gg 1} =
\frac{r_S}{c\,u^{\,2}}$. The second and main phase delay is the
gravitational redshift (Shapiro Phase Delay) due to a MACHO field
$\sim \int \frac{G M}{c^2 r} dr$ all along the wave trajectory;
its behaviour is logarithmic on the impact parameter $u$:\\
\begin{equation}\label{eq1}
\Delta t_{grav} = \frac{r_S}{c} \ln \left( \frac{8\,D_s}{r_S}
\right) \,-\, \frac{2\,r_S}{c} \ln \left(\, \sqrt{u^2 + 4 \;}
\,+\, u\, \right)
\end{equation}
\smallskip
where $u = b \,/\, R_E$ is the adimensional impact parameter of
any MACHO crossing along, whose Schwarchild and Einstein radii
are:\\
\[
r_S = \frac{G M}{c^2} \;\;\;\;,
\]
\begin{equation}\label{eq2}
R_E = \sqrt{2 r_S \frac{D_d D_{ds}}{D_s}} = 4.8 \cdot 10^{13}
\sqrt{\left[\frac{M}{M_\odot}\right] \left[\frac{D_s}{4\,
Kpc}\right]\left[\frac{x_d}{1/2}\right]\left[2-\frac{x_d}{1/2}\right]}
\;\;\; cm
\end{equation}
\smallskip
where $x_d = \frac{D_d}{D_s}$, $1 - x_d = \frac{D_{ds}}{D_s}$;
$D_d$ is the deflector-observer distance, $D_{ds}$ the
source-deflector distance and $D_s$ the source-observer
distance.\\ The consequent characteristic time of the Shapiro
delay while the MACHO approaches or goes far away the line of
sight is:\\
\begin{equation}\label{eq3}
t_c = \frac{R_E}{v_\perp} u_{min} \simeq 4.55 \;
\frac{\left[\frac{u_{min}}{10^2}\right]}{\left[\frac{\beta_\perp}{10^{\,-3}}\right]}\,
\sqrt{\left[\frac{M}{M_\odot}\right] \left[\frac{D_s}{4\,
Kpc}\right]\left[\frac{x_d}{1/2}\right]\left[2-\frac{x_d}{1/2}\right]}
\;\;\; yr
\end{equation}
\smallskip
and the consequent PSR's period derivative is:\\
\begin{equation}\label{eq4}
\dot{P} = 1.38 \cdot 10^{\,-13} \;
\frac{\left[\frac{\beta_\perp}{10^{\,-3}}\right]\,\sqrt{\frac{M}{M_\odot}}}{\left[\frac{u_{min}}{10^2}\right]
\sqrt{ \left[\frac{D_s}{4\,
Kpc}\right]\left[\frac{x_d}{0.5}\right]\left[2-\frac{x_d}{0.5}\right]}}
\;\;\;s\,s^{\,-1}
\end{equation}
\smallskip
First we must notice that the above period derivative is an order
of magnitude larger than the average PSR one ($\dot{P} \sim
10^{\,-14}\;s\,s^{\,-1}$) and therefore it is well detectable.
Secondly because of above sensitivities we could parametrized the
adimensional impact parameters not just as $u_{min} \sim 1$ as
needed in microlens technique but at $u_{min} \sim 10^2$ scales.
This offer a corresponding square geometrical probability
amplification ($\pi u^2$) to reveal a MACHO, nearly  four orders
of magnitude larger than in microlensing case. For this reason
present Shapiro Delay on 7 hundreds PSRs  are comparable with a
sample of nearly 7 millions of stars for microlensing. However a
positive period derivative ($\dot{P} > 0$) might be well indebted
to a large intrinsic positive angular momentum loss. Therefore we
first looked for negative PSR period derivative ($\dot{P} < 0$)
whose interpretation cannot be indebted to any (rare) corotating
accretion mass (single stars). We could find statistically a few
events a year [5], for a MACHO density comparable with the
observed microlenses. We did observed (up date to 6) and at least
one, B1813-26, very isolated PSR, which we do definitively
interpret as a Shapiro Phase Delay alive. We also find a group of
PSRs in a globular cluster possibly suffering a collective Shapiro
Phase Delay due to the own cluster gravitational field: B0021-72C,
B0021-72D in 47 Tucanae and B2127+11D, B2127+11A in M15.

\bigskip
\section{Shapiro Delay in Dark Planet search}
\bigskip

The technique above could be fruitful and certain believed by
testing and calibrating the PRS's delay while crossing along the
Sun's field, Juppiter or Mars trajectories. For nearby deflectors
the gravitational phase delay has a more complicated 3D vectorial
formula:\\
\begin{equation}\label{eq5}
\Delta \; t_{Grav} \;=\; \frac{r_S}{c} \; \ln \left[ \; \frac{ \,
D_s \,-\, D_d \, \cos \psi \,+\,D_{ds}\;}{D_d \, \left(\, 1 \,-\,
\cos \psi \,\right) } \;\right]
\end{equation}
\smallskip
where $D_d = |\overrightarrow{D_d}|$, $D_{s} =
|\overrightarrow{D_{s}}|$, $D_{ds} = |\overrightarrow{D_{s}} -
\overrightarrow{D_{d}}| = |\overrightarrow{D_{ds}}|$ and $\psi =
\widehat{\overrightarrow{D_{s}} \overrightarrow{D_{d}}}$. For
$\psi \rightarrow 0$ after simple expansion one recovers from
equation \ref{eq5} the previous Shapiro well known equation
\ref{eq1}. For the Sun the effect even few degrees far away its
position is huge ($|\dot{P}| \sim 10^{-10}\;s\,s^{\,-1}$) and must
be observed within $t_c \sim day$. In order to avoid any
additional and confusing refractive index delay (due to solar
plasma even at large impact parameters) PRSs might be better
observed near Sun at higher radio PSR frequencies [5] ($\nu \geq
GHz$) where plasma refractive index is smaller than the
gravitational Shapiro one. Solar Shapiro delay has, naturally, a
strong annual modulation. Planets like Jupiter, at $5.2\,AU$, will
produce well detectable period derivative ($|\dot{P}| \sim
10^{-12}\;s\,s^{\,-1}$) at $u_{min} \sim 100$ and even at
$10^\circ$ angular impact parameter distance the effect might be
observable ($|\dot{P}| \geq 6 \cdot 10^{-15}\;s\,s^{\,-1}$) (see
figure 2). In particular to test the technique we chose a sub
sample of all known PRSs laying along the ecliptic trajectory
($\pm 6^\circ$). Most PRSs are localized toward Galactic Center
and Anticenter and are described in figure 3 [5]. The planetary
(Juppiter, Mars..) Shapiro Delay is strongly modulated not only by
the Terrestrial yearly trajectory, but also by the combined
Earth-Juppiter, Earth-Mars.. mutual distances, which play a key
role in defining the Einstein radius and the Shapiro Phase Delay.
These combined effects must introduce a very characteristic
imprint in their own phase delay as we predicted in figure 3.

\bigskip
\section{ Conclusions}
\bigskip

The Shapiro Phase might be able to discover not only dark matter
but once calibrated it may even lead to new mini planets discovers
on our ecliptic plane. Their presence might be of relevance (by
tidal disturbances during rare encounter) on Earth past (and
future) history (and life evolution \cite{dfdar}). Finally recent
EGRET data \cite{dixon} on diffused Gev galactic-Halo might find
an answer, among the others \cite{df3}, by ($\sim pc$) hydrogen
molecular clouds interacting by proton (Gev) cosmic rays
\cite{depaolis}. Microlenses at those large radii, (by Kirchoff
theorem) are unefficient. Therefore the Shapiro Phase Delay might
be the unique probe to verify such an evanescent dark matter
barionic candidature.

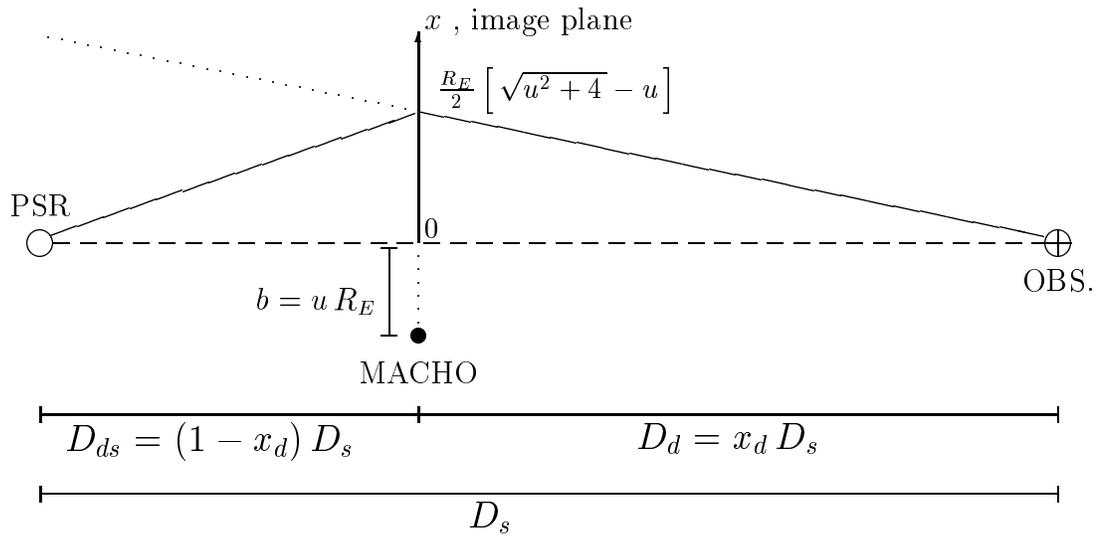
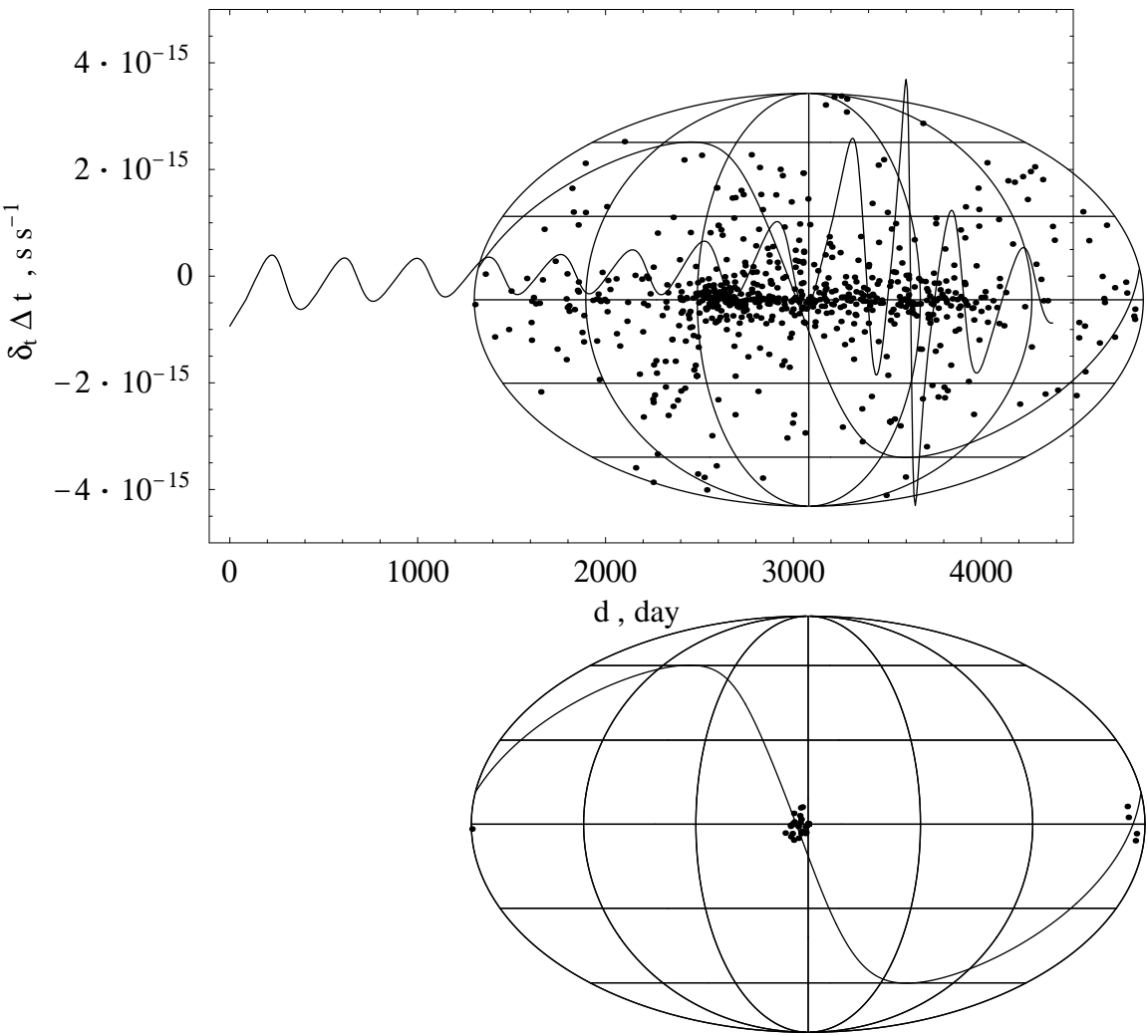
\end{document}